\documentclass[preprint,preprintnumbers,amsmath,amssymb]{revtex4}
\usepackage{graphicx}

\begin{document}

\title{Prospects for detection of the lunar Cerenkov emission 
by the UHE Cosmic Rays and Neutrinos using the GMRT 
and the Ooty Radio Telescope}

\author{Govind Swarup$^{a}$ and Sukanta Panda$^{b,c}$}%
\affiliation{$^a$National Centre for Radio Astrophysics-TIFR, Pune 411007\\
$^b$Departamento de Fisica Teorica C-XI, 
Universidad Aut{\`o}noma de Madrid, Cantoblanco, 28049 Madrid, Spain\\
$^c$Instituto de Fisica Teorica UAM/CSIC,
Universidad Aut{\`o}noma de Madrid, Cantoblanco, 28049 Madrid, Spain}

\begin{abstract}
Searching for the Ultra high energy Cosmic rays and 
Neutrinos of $> 10^{20} eV$ is of great cosmological importance. 
A powerful technique is to search for the \v{C}erenkov radio emission 
caused by UHECR or UHE neutrinos impinging on the lunar regolith. 
We examine in this paper feasibility of detecting these events by 
observing with the Giant Metrewave Radio Telescope (GMRT) 
which has a large collecting area
and operates over a wide frequency range with an 
orthogonal polarisation capability. We discuss here prospects
of observations of the \v{C}erenkov radio emission with the GMRT
at 140 MHZ with 32 MHz bandwidth using the incoherent array
and also forming 25 beams of the Central Array 
(effective collecting area of 14250 $m^2$) to cover the moon.
We also consider using the Ooty Radio Telescope (ORT) which
was specially designed in 1970 for tracking the Moon.
The ORT consists of a 530m long and 
30m wide parabolic cylinder that is placed in the north south 
direction on a hill with the same slope as the latitude of the station. 
Thus it becomes possible to track the Moon for 9.5 hours on a 
given day by a simple rotation along the long axis of the 
parabolic cylinder. ORT operates at 325 MHz and has an 
effective collecting area of $~$ 8000 $m^2.$ Recently a digital system 
has been installed by scientists of the Raman Research Institute (RRI), 
Bangalore and the Radio Astronomy Centre (RAC) of NCRA/TIFR, at 
Ooty allowing a bandwidth of 10 MHz with $~$ 40 ns sampling. It is 
possible to form 6 beams covering the Moon and 7th beam far away for 
discrimination of any terrestrial RFI. Increasing the 
bandwidth of the existing 12 beam analogue system of the ORT 
from 4 MHz to 15 MHz to be sampled digitally is planned. It is shown 
that by observing the Moon for $\ge$ 1000 hrs using 
the ORT it will provide appreciably higher sensitivity 
than past searches made elsewhere 
and also compared to the search being made 
currently in Netherlands using the Westerbork 
Synthesis Radio Telescope (WSRT) at 140 MHz. 
Using the GMRT and ORT, it may be possible to reach 
sensitivity to test the Waxman-Bachall limit based on 
theoretical arguments on the UHE particle flux. 
\end{abstract}

\maketitle

\section{Introduction}
Several theoretical predictions and scenarios have been proposed 
for the existence of ultra high energy (UHE) cosmic rays (UHECR) 
and UHE neutrinos $> 10^{20} eV.$ Detection of the UHECR and 
particularly UHE neutrinos would be of great importance 
for understanding the energy of powerful AGNs, 
gamma ray bursts and possible existence of massive particles 
predicted by the GUT theories. For detecting UHECR and 
UHEN several ambitious terrestrial experiments are 
being carried out and also planned with very large 
collecting areas $>$ 1 $km^{2}$ and volumes $>$ 1 $km^{3}$~\cite{arianna}. 

Askaryan noted in 1960s~ \cite{aska}, that electromagnetic cascades 
in dense medium by the UHE particles will develop an 
excess of negative charge giving rise to incoherent \v{C}erenkov 
radiation. Later, Dagkesamanskiĭ and Zheleznykh~ \cite{dz} noted 
that UHE particles impinging on the lunar 
regolith at $\sim$ 10 m-20 m deep layers of the Moon will 
give rise to radio pulses of nanosecond (ns) durations. 
The large surface area of the Moon effectively provides a 
large surface area for detection of the rare UHE particles. 
Observations have been made towards the Moon at 1.4 GHz 
using the Parkes 64 m diameter radio telescope~ \cite{parkes}, 
and at 2.2 GHz using the JPL/NASA 70 m and 30m 
antennas (GLUE experiment)~ \cite{glue} 
and using a single 64 m telescope at 
Kalyazin Radio Astronomical Observatory~ \cite{kalyazin}. 
These have put upper limits on the existence of UHE particles 
but these are appreciably higher than the 
predictions by Waxman and Bahcall~\cite{Waxman:1998yy}. 
Askaryan effect has been tested using different media 
in a series of accelerator experiments. One of such experiment
is done in Silica sand which resembles 
composition of lunar regolith~\cite{slac1}.
 
As shown by Alvarez-Muniz et al.~\cite{ZHS}, the angular distribution of 
the electric field emitted by 10 Tev shower in ice, salt and the 
lunar regolith is much wider at 0.1 GHz than at ~ 1 GHz. 
Scholten et al.~\cite{Scholten} have calculated differential detection 
probability for cosmic rays of energy $4\times 10^{21}$ eV and 
neutrinos of energy $2\times 10^{22}$ eV hitting the Moon 
as a function of apparent distance from the centre of the 
Moon for different detection frequencies. It is shown that 
the radio emission at higher frequencies arises mostly from 
UHE particles impinging near the rim of the Moon but 
at lower frequencies from a major part of the Moon, 
indicating the advantage of making observations at lower 
frequencies using already existing or planned 
radio telescopes of large collecting areas in the frequency
range of about 30 to 300 MHz. For detecting UHECR and UHE neutrinos, 
observations are currently being carried out by radio 
astronomers in Netherlands using the Westerbork 
Radio Telescope (WSRT)~\cite{bruyn} at $~$ 140 MHz. Observations are 
also planned with the LOFAR~\cite{bregman} under construction.

In Section II, we summarize equations giving the 
expected value of the electric field and flux density for 
UHE particles as well as 25 times rms detection threshold 
of a radio telescope of collecting area $A_{eff}.$ 
Panda {\it et al}~\cite{panda} have recently considered 
prospects of using the Giant Metrewave Radio 
Telescope (GMRT)~\cite{swarup} for observing 
radio pulse emission arising from the UHE particles 
interacting with the surface of the Moon. In Section III, we 
describe appropriate parameters of the GMRT for searching 
the lunar \v{C}erenkov emission and also summarize expected 
values of the signal strength as a function of energy of 
UHE particles and the receiver noise threshold. 

In Section IV, we propose observations of the \v{C}erenkov radiation 
from the lunar regolith using the 
large Ooty Radio Telescope (ORT)~\cite{swarup1} 
that has an effective collecting area, $A_{eff}$ = 8000 $m^2$ 
and is operating at 325 MHz. At present ORT provides a 
bandwidth of only 4 MHz but its receiver system has been modified 
to provide $\Delta \nu = 10$ MHz~\cite{prabhu}  and 
is being extended to 15 MHz. 
In contrast to the GMRT providing dual polarizations 
at several frequency bands, the ORT provides only 
a single polarization but it would be 
possible to get observing time of $> 1000$ hours, as it is 
being used mostly for day time 
interplanetary scintillations.

As discussed in Sections IV and V, search for UHE particles 
will also allow simultaneous observations of lunar 
occultation of radio sources in the path of the Moon 
and also variation of brightness temperature of the Moon 
with the lunar phase, the latter yielding parameters such as
dielectric constant and electrical conductivity of  
the lunar regolith upto depths of 30 m to 100 m.

In Section VI we discuss model independent limits for 
detection of UHECR and UHE neutrinos for several 
current and planned experiments, including LOFAR, WSRT, 
GMRT and ORT. Discussions and Conclusions 
are given in Section VII.

\section{Estimated strength of radio waves 
from cascades in the lunar regolith}                         

The electric field of radio waves on Earth, ${\cal E},$ from 
a \v{C}erenkov shower in the lunar regolith due to 
UHE neutrinos, with energy $E_s,$ has been 
parameterized based on accelerator measurements 
and Monte Carlo simulations ~ \cite{ZHS, Scholten}
(neglecting angular dependence) giving
\begin{equation}
{\cal E}(V m^{-1} MHz^{-1})= \left(\frac{2.53\times 10^{-7}}{R}\right) \, 
\left(\frac{E_s}{\mathrm{TeV}}\right) \times 
\left(\frac{\nu}{\nu_0[1+(\nu/\nu_0)^{1.44}]}\right)\, .
\label{field}
\end{equation}
where R is the distance between the emission point on 
the Moon's surface to the telescope, $\nu$ is the 
radio frequency of observations and $\nu_0=2.5$ GHz for the 
lunar regolith material. 

The power flux density at Earth, $F_s$ is given by
\begin{equation}
F_s = ({\cal E}^2/Z_0) (\Delta \nu/ 100 MHz) \times 10^{22} Jy.
\label{intensity}
\end{equation}
where free space impedance, $Z_0$ = 377 ohms, receiver bandwidth, 
$\Delta \nu,$ is in units of 100 MHz and 1 {\it Jy} = $10^{-26} W m^{-2} s^{-1}.$

Substituting from Eq. \ref{field}, we get
\begin{equation}
F_s = 1.7 \times 10^6 \left(\frac{1}{R}\right)^2  
\left(\frac{E_s}{1TeV}\right)^2 
\times
\left(\frac{\nu}{\nu_0[1+(\nu/\nu_0)^{1.44}]}\right)^2\, 
(\Delta \nu / 100 MHz) Jy.   
\end{equation}
Panda et al.~\cite{panda} has given the following 
value of the power flux density
\begin{equation}
F(E_s,\theta,\nu)= 5.3 \times 10^{5} \, f(E_s,\theta) 
\left(\frac{E_s}{\mathrm{TeV}}\right)^2 \times
\left(\frac{1\,\mathrm{m}}{R} \right)^2 
\left(\frac{\nu}{\nu_0[1+(\nu/\nu_0)^{1.44}]}\right)^2 \, 
\frac{\Delta \nu}{100\,\mathrm{MHz}}\,\,\, \mathrm{Jy}.
\label{F}
\end{equation}
Furthermore there is an angular dependence given by
\begin{equation}
 f(E_s,\theta)=\left(\frac{\sin\theta}{\sin\theta_C}\right)^2\exp(-2Z^2)
\label{f}
\end{equation}
 with $Z= \sqrt{\kappa} (\cos\theta_C-\cos\theta)$ and 
$\kappa(E_s)=(\nu/\mathrm{GHz})^2(70.4 + 3.4 \ln(E_s/\mathrm{TeV})).$ 
Here we used Gaussian approximation for our calculation where 
the forward-suppression factor $\sin^2\theta$ in (\ref{f}) is ignored. 
For high frequencies this has no effect. For low frequencies, the 
differences at small angles only plays a role for showers nearly 
parallel to the surface normal, while the effects of changing 
the normalization near the \v{C}erenkov angle is important 
also for more horizontal showers. A measure of the effective 
angular spread $\Delta_C$ of the emission around 
the \v{C}erenkov angle $\theta_C$ is given in terms of
\begin{equation}
\Delta_C=\sqrt{\frac{\ln(2)}{\kappa(E_s)}}\frac{1}{\sin\theta_C}.
\end{equation}

 It is seen from above that the value of $F_s$ as given 
by Panda et al. is about 3 times lower than that given 
by Eq.(2). We find that the value of F by 
Panda {\it et al}~\cite{panda} is ~ 0.92 that given 
by Scholten {\it et al}~\cite{Scholten}. We have used here Eq.\ref{F} 
as per Panda {\it et al}~\cite{panda}.

By equating the power, $\Delta p,$ received by a radio telescope, 
due to the incident input threshold threshold power flux density, 
$F_s=F_N,$ with the minimum detectable receiver noise power, 
$k \Delta T \Delta \nu,$ we have 
$\Delta p = \frac{1}{2}F_N A_{eff} \Delta \nu,$ 
where the factor $1/2$ is due to the reception of a single
polarization, $A_{eff}$ the effective area of the telescope,
$\Delta \nu$ the bandwidth and the receiver rms noise, 
$\Delta T=\frac{k_B T_{s}}{\sqrt{\Delta t \Delta \nu}},$ $T_s$
being the system temperature, $k_B$ Boltzmann's constant and
$\Delta t$ the integration time. Hence $F_N$ is given by 
\begin{equation}
F_N= \frac{2 k_B T_{s}}{\sqrt{\Delta t \Delta \nu} A_\mathrm{eff}}.
\end{equation}                                                                                                                            
For detection of a narrow pulse with width $\Delta t$ using 
an optimum bandwidth $\Delta \nu,$ 
$(\Delta \nu \Delta t)^{1/2}\approx 1,$ and hence rms noise $F_N$ is given by
\begin{equation}
F_N = \frac{2 k_B T_{sys}}{A_\mathrm{eff}}.                         
\end{equation}                                                                                           
In Tables \ref{table1}, \ref{table2}, \ref{table3} we list the system 
temperatures at the different observation
frequencies and the corresponding noise levels of two different 
configurations of GMRT and ORT. Using equation (\ref{F}), we can 
solve for $E_s$ at the threshold required for measurement with 
the radio telescope (obtained for $\theta=\theta_C$ and $F=F_N$). If 
we take a required signal-to-noise 
ratio $\sigma,$ the threshold shower 
energies $E_\mathrm{th}$ which can be 
measured at the different 
observation frequencies at the GMRT and the ORT are given in 
Tables \ref{table1}, \ref{table2}, \ref{table3}.

\section{Prospects for searching using the 
Giant Metrewave Radio Telescope (GMRT)}

The GMRT is a Synthesis Radio Telescope consisting 
of 30 nos. of fully steerable parabolic dish antennas 
each of 45 m diameter. Fourteen antennas are 
located in a somewhat random array within an area 
of about 1 $km^2$ and other sixteen antennas along 
3 Y-shaped arrays with a length of each $\sim$ 14 km. 
The GMRT is currently operating in 5 frequency bands 
ranging from about 130 MHz to 1430 MHz. The receiver 
system provides output at two orthogonal 
polarizations from each of the 30 antennas with a maximum bandwidth 
of 16 MHz for each polarization, being sampled at 
32 ns each. The $A_{eff}$ of each 
antenna is nearly 950 $m^2$ in the frequency range 
of 130 to 630 MHz and only 600 $m^2$ at 1430 MHz.

Panda et al. have made estimates of the sensitivity of the GMRT 
for observations of UHE CR and UHE neutrinos. 
They have considered the $A_{eff}$ 
of the GMRT $=30,000 m^2$ at 150, 235, 325 and 610 MHz and 
18,000 $m^2$ at 1390 MHz. However, we may note 
that the GMRT provides the above area only 
when the voltage outputs of all the 30 antennas are 
added in phase resulting in antenna beam of $\sim$ 2 arcsec at 
the highest frequency and $\sim$ 15 arcsec at 150 MHz and therefore
covering only a small part of the Moon. 
However the receiver correlator allows 
incoherent addition of the outputs of the 
30 antennas, covering the entire front surface
of the Moon and resulting in $A_{eff-incoherent}$ 
$\sim 30^{1/2} \times 950 = 5203 m^2$ at the lower 
4 frequency bands and $30^{1/2} \times 600 = 3286 m^2$ 
at 1390 MHz. Insetad if we measure coincidences of the 
power outputs of the 30 antennas, the effective area will also 
be 5203 $m^2$ at the lower frequency bands but would have
the advantage of discrimination between the lunar Cerenkov emission
and any terrestrial radio frequency interference(RFI) as the GMRT
antennas are located in an array of $\sim$ 25 km extent.

An alternative strategy will be more effective if we 
use the recently installed software correlator at the GMRT for 
cross multiplications of the voltage outputs of the 30 GMRT 
dishes with $\Delta \nu=16$ MHz. 
It allows 32 ns sampling of the voltage outputs of 
each of the 30 antennas. By combining these voltage 
outputs for the central 14 antennas of the GMRT 
with appropriate phase values, it would be possible to 
form 25 phased beams covering the Moon, each beam having 
a resolution of about 6 arcmin at 140 MHz. 
The effective area for each of the 25 beams will be 14250 $m^2$ 
at the lower frequency bands and 9000 $m^2$ at $\sim$ 1 GHz, 
providing a competitive radio telescope 
for searching for UHE neutrinos. Contributions by the 
Moon's temperature to the 
system temperature of the GMRT receiver is 
negligible at 140 MHz but is appreciable at higher frequencies. 
Using the system parameters of the GMRT as 
given in Tables \ref{table1} and \ref{table2}, 
we have estimated sensitivity of 
UHE CR and UHE neutrinos fluxes as given in Figs  
\ref{crlimit100}, \ref{crblimit100}, \ref{nlimit100}, 
\ref{nblimit100} and \ref{fluxgmrt}. 
\begin{table}[htp]
\begin{tabular}{|l|l|l|l|l|l|l|l|l|r|}
\hline
$\nu$ (MHz) & $\Theta_b$ (deg) & $(T_m - T_{sky})$ (K)
& $T_s$ (K) & $\Delta \nu$ (MHz) & $F_N$ (Jy) &
${\cal E}_N \,$($\mu$V/m/MHz) & $E_\mathrm{th}(0)/\sqrt{\sigma}$ (eV)
\\ \hline
  140 & 3.0 & 0 & 450 & 16 & 238 & .0075 & $3.56 \times 10^{20}$ \\
  235 & 1.9 & 150 & 187 & 16 & 99 & .0048 & $1.4 \times 10^{20}$ \\
  325 & 1.4 & 190 & 132 & 16 & 70 & .004 & $8.62 \times 10^{19}$ \\
  610 & 0.73 & 220 & 204 & 16 & 108 & .005 & $ 6.13 \times 10^{19}$ \\
  1390& 0.32 & 225 & 297 & 16 & 249 & .0076 & $ 5.16 \times 10^{19}$ \\
\hline
\end{tabular}
\caption{GMRT parameters, sensitivity and 
threshold sensitivity at different frequencies for an 
incoherent array. $\Theta_b$ is the 
full width half maximum(FWHM) beam of the 45m dishes, 
$T_m$ is the temperature 
of the Moon at frequency $\nu.$ 
$F_N$ is the expected threshold flux 
density (noise intensity) of the GMRT and ${\cal E}_N$ the 
corresponding electric field. The threshold energy 
$E_\mathrm{th}$ is given in the last column.}
\label{table1}
\end{table}

\begin{table}[htp]
\begin{tabular}{|l|l|l|l|l|l|l|l|l|r|}
\hline
$\nu$ (MHz) & $\Theta_b$ (deg) & $(T_m - T_{sky})$ (K)
& $T_s$ (K) & $\Delta \nu$ (MHz) & $F_N$ (Jy) &
${\cal E}_N \,$($\mu$V/m/MHz) & $E_\mathrm{th}(0)/\sqrt{\sigma}$ (eV)
\\ \hline
  140 & 3.0 & 0 & 450 & 16 & 87 & .0045 & $2.14 \times 10^{20}$ \\
  235 & 1.9 & 150 & 187 & 16 & 36 & .0029 & $8.38 \times 10^{19}$ \\
  325 & 1.4 & 190 & 132 & 16 & 25 & .0024 & $5.15 \times 10^{19}$ \\
  610 & 0.73 & 220 & 204 & 16 & 39 & .003 & $3.68 \times 10^{19}$ \\
  1390& 0.32 & 225 & 297 & 16 & 91 & .0046 & $ 3.12 \times 10^{19}$ \\
\hline
\end{tabular}
\caption{GMRT parameters, sensitivity and
threshold sensitivity at different frequencies 
for 25 beams case. $\Theta_b$ is the
full width half maximum (FWHM) beam of the 45m dishes, 
$T_m$ is the temperature
of the moon at frequency $\nu$. $F_N$ is the expected noise
intensity of GMRT and ${\cal E}_N$ the corresponding electric field.
The threshold energy $E_\mathrm{th}$ is given in the last 
column.}
\label{table2}
\end{table}

\begin{table}[htp]
\begin{tabular}{|l|l|l|l|l|l|l|r|}
\hline
$\nu$ (MHz) & $T_s$ (K) & $\Delta \nu$ (MHz) & $F_N$ (Jy) &
${\cal E}_N \,$($\mu$V/m/MHz) & $E_\mathrm{th}(0)/\sqrt{\sigma}$ (eV)
\\ \hline
  325  & 200 & 15 & 67.5 & .0041 & $8.74 \times 10^{19}$ \\
\hline
\end{tabular}
\caption{ORT parameters, sensitivity and
threshold sensitivity. $F_N$ is the expected noise
intensity of ORT and ${\cal E}_N$ the corresponding electric field.
The threshold energy $E_\mathrm{th}$ is given in the last
column.}
\label{table3}
\end{table}
\section{Prospects for searching UHE CR and neutrinos using
the Ooty Radio Telescope(ORT)}

The ORT consists of a 530m long and 30m wide
parabolic cylinder that is placed in the north
south direction on a hill with the same slope
as the latitude of the station~\cite{swarup1}.
Thus it becomes possible to track the Moon for
9.5 hours on a given day by rotating the parabolic 
cylinder along it's long axis. The ORT operates only
at 325 MHz and has effective collecting area of
$\sim 8000 m^2.$ A phased array of 1056 dipoles is
placed along the focal line of the parabolic
cylinder. Each dipole is connected to an RF
amplifier followed by a 4 bit phase shifter. 
Signals received by 48 dipole units are 
connected to a common amplifier branching 
network ~\cite{sel}. The 22 outputs of the
phased array are brought to a central receiver room.
An analogue system that was originally built for
lunar occultation observations~\cite{swarup1} provided
12 beams to cover the Moon; each beam is 6 arcmin
in the north side direction and 126 arcmin ($\sim$ 2 deg.) 
in the east west direction. Recently a digital
system has been installed by the Raman Research
Institute (RRI), Bangalore and the Radio Astronomy
Centre of NCRA/TIFR, at Ooty allowing formation
of phased array beams with collecting area = 8000 $m^2$
and a bandwidth of 10 MHz with $\sim$ 40 ns sampling~\cite{prabhu}.
It is possible to form 6 beams covering
the Moon and 7th beam far away for discrimination
of any terrestrial RFI. The proposed upgrade
of the 12 beam analogue system will provide 
a bandwidth of 15 MHz. The measured receiver
temperature of the ORT is 140 K + a contribution by
moon of $~$ (31.5/126) x 230 K = 57 K. Thus $T_{s}$ of ORT
for lunar observations at 327 MHz is about 200 K.
As discussed in the next Section, observations
of the Moon for 1000 hrs using the ORT at
325 MHz will provide appreciably higher sensitivity
than the past searches made by various workers
and also compared to a search being made currently
in Netherlands using the Westerbork Synthesis
Radio Telescope (WSRT) at 140 MHz. Using the ORT,
it may be possible to reach sensitivity to test
the predictions of the Waxmann-Bachall model
based on theoretical arguments. Proposed observations, particularly
with the ORT will also provide arcsec resolution for galactic and
extragalactic radio sources occulted by the Moon, and may also
search for any transient celestial sources in the antenna beam
outside the disc of the Moon.

\section{Measurement of dielectric constant and 
electrical conductivity of the lunar regolith}

It would be quite valuable to make passive radio maps of 
the Moon using the GMRT at decimetre and metre wavelengths. 
The suface temperature of the Moon is about 130 K 
in its night time and $\sim$ 330 K in its day time. 
Since Moon's surface consists of lossy dielectric material, 
the radio waves emitted by its thermal properties 
arise from few cm at microwaves to more than 100 m 
deep at wavelength of several m. Therefore, the observed 
values of brightness temperature of the Moon varies 
by tens of degrees at microwaves to less than a degree 
at radio wavelengths. The GMRT provides a resolution of 
about 2 arcsec at $\sim$ 1420 MHz and $\sim$ 15 arcsec at 150 MHz. 
Polarization observations are also possible with 
the GMRT. Therefore, maps of radio emission of the 
Moon for its night and day with the GMRT will 
provide estimates of the dielectric constant and 
electric conductivity of the lunar regolith. 
The data will be complimentary to the radar measurements~\cite{radar}. 

\begin{figure}[t]
\hbox{\hspace{0cm}
\hbox{\includegraphics[scale=1]{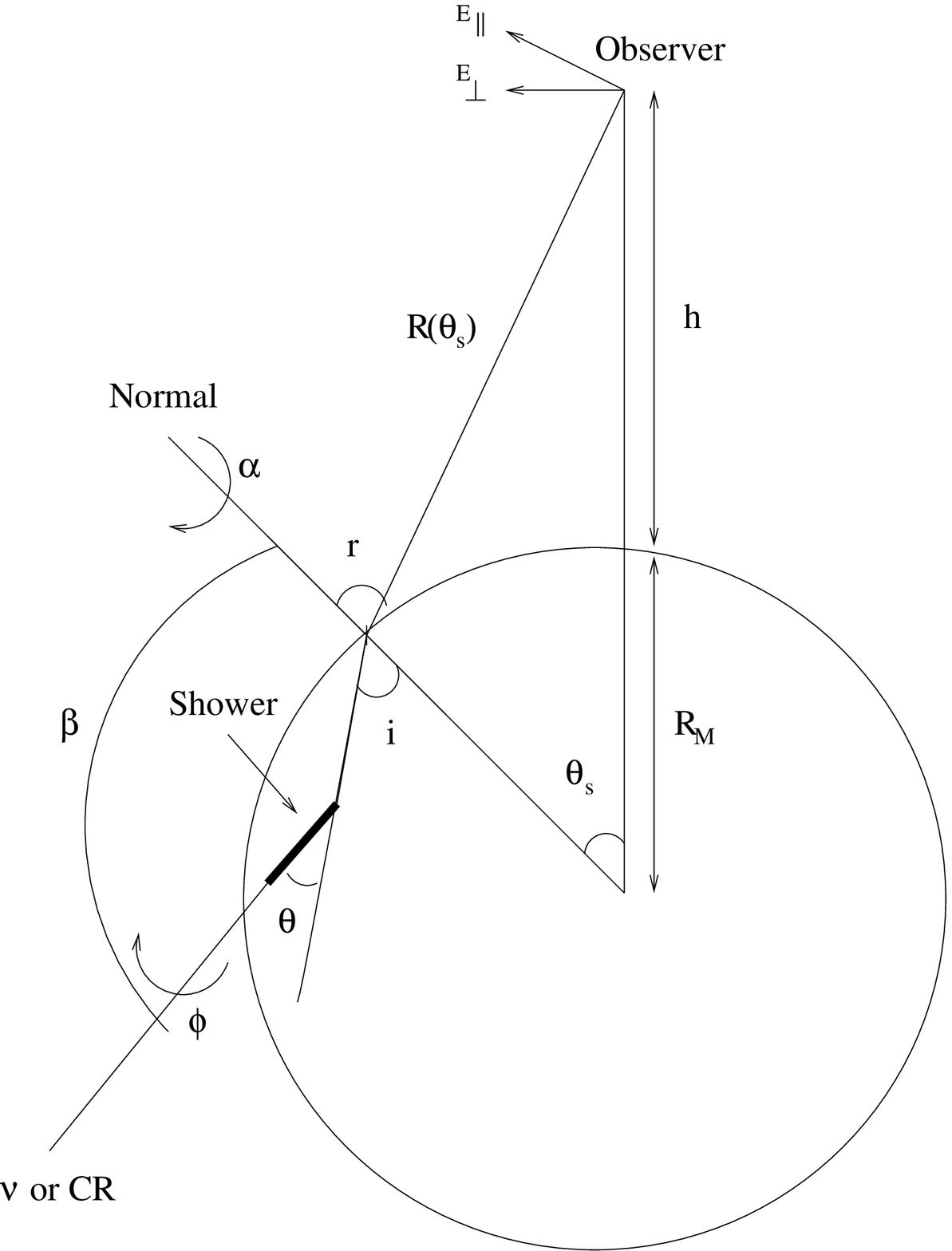}}}
\caption{Geometry of an UHE CR or neutrino event which generates 
\v{C}erenkov radiation of radio waves in the lunar regolith. 
The Earth is located at a mean distance $h$ from 
the lunar surface \cite{panda}.}
\label{geometry}
\end{figure}
\section{Sensitivity calculations for search of 
UHE neutrinos and Discussions}

In this Section we calculate 
model independent limits for detection of UHE CR 
and UHE neutrinos for GMRT and ORT using the procedure
given in Panda {\it et al} \cite{panda}.

Scholten et al.~\cite{Scholten} have considered 
dividing the WSRT antennas into 3 groups for 
the proposed search for neutrinos, whence A is 
likely to be $~ 1000 m^2.$ They give a value of $F_N$ = 600 Jy 
for WSRT and $F > 25 F_N = 15000$ Jy. The system temperature 
for the ORT, $T_{s}$ = 200K, at 327 MHz 
including contribution by the Moon and 
$A_{eff}=8000 m^2$. Hence $F_N = 67.5 Jy$ and $F > 25 F_N$ = 1687.5 Jy, 
which are much lower than for the WSRT. 

The event rate that would be expected at the telescope 
can be related to an isotropic flux $\Phi$ of 
UHE particles on the Moon through
\begin{equation}
\frac{\mathrm{d}N_i}{\mathrm{d}t}= \int \frac{\mathrm{d}\Phi_i}
{\mathrm{d}E} A_i
(E) \mathrm{d}E,
\end{equation}
where $i=\{\mathrm{CR},\nu\}$ denote the type of 
primary particle and $A_i(E)$ is an aperture function 
corresponding to the effective detector area. 
The aperture can be 
further decomposed into an angular aperture 
$\Delta\Omega_i(E,\theta_s)$ and a geometric area 
factor for the Moon
\begin{equation}
A_{i}(E)= 2 \pi R_{M}^2 \int \Delta\Omega_{i}(E,\theta_s)
\mathrm{d}\cos\theta_s
\label{acr}
\end{equation}
with $R_M=1760$ km. To evaluate the aperture, we use 
the analytical methods described in~ \cite{gusev}. 
For the case of strongly interacting cosmic rays 
which can mainly interact on the surface of the Moon, 
the angular aperture is given by
\begin{equation}
\Delta\Omega_{CR}(E,\theta_s)=\int \cos\beta
\Theta[{\cal E} (E,\theta_s)-{\cal E}_{th}] \times \Theta(\cos\beta)
\mathrm{d}\alpha\mathrm{d}\cos\beta,
\label{angle}
\end{equation}
where $\beta$ and $\alpha$ are the polar and 
azimuthal coordinates of the ray normal to 
the Moon's surface in a system where the 
shower direction defines the $z$ axis. The full 
geometry and the different angles are 
described in Fig.~\ref{geometry}.

When the UHE primary is instead a neutrino, it 
can produce showers deep below the surface of 
the Moon and there will be considerable attenuation 
of the radio waves which travel distances longer 
than $\lambda_r$ below the surface. For the 
neutrino induced showers, the aperture is defined 
in the same way as for the CR, 
but the angular aperture is now given~ \cite{gusev} by
\begin{equation}
\Delta\Omega_{\nu}(E,\theta_s)=\int 
\frac{\mathrm{d}z}{\lambda_{\nu}}
\int\Theta\bigl\{{\cal E}(E,\theta_s, \theta)
\exp[-z/(2 \lambda_r \cos i)]-{\cal E}_{th}\bigr\} 
\times \exp[-L(z,\beta)/\lambda_{\nu}]
\times \mathrm{d}\alpha \mathrm{d}\cos\beta,
\label{omega-nu}
\end{equation}
where $L(z,\beta)$ is the distance the neutrino travels
inside the material to reach the interaction point at a
distance $z$ below the surface. In performing this 
integration we allow $z$ to go below the known 
depth of the regolith. Despite the attenuation,
the aperture therefore picks up contributions coming
from deep showers, especially for the lower 
frequencies. Numerically we find for the worst 
case (when $\nu=150$ MHz), that imposing
a sharp cutoff at a depth of $20$~m would reduce the
aperture by nearly an order of magnitude, 
similarly to what was discussed in~ \cite{Scholten}.

As for the cosmic rays, the total aperture is obtained
by substituting (\ref{omega-nu}) into (\ref{acr}) and 
integrating over the polar angle $\theta_s$.

To estimate the sensitivity of GMRT and ORT to cosmic ray and 
neutrino events we have evaluated the angular apertures 
by employing this technique and performing numerical 
integrations for the different parameters given 
in Tables \ref{table1},\ref{table2}, \ref{table3}. 
In the next section we will discuss 
these results further in the context of prospectiveflux limits.

\begin{figure}[t]
\hbox{\hspace{0cm}
\hbox{\includegraphics[scale=1.3]{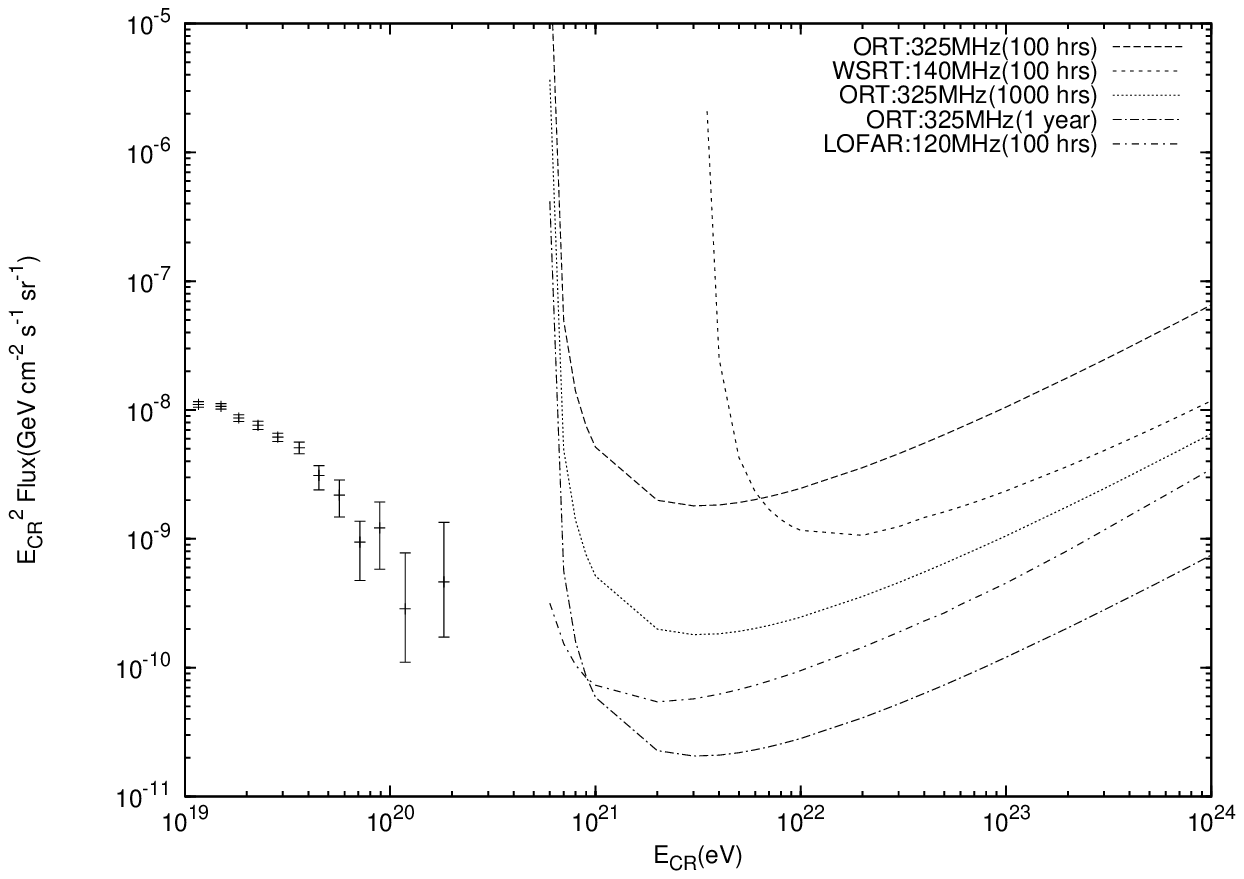}}}
\caption{Model independent limits on UHECR flux at different
frequencies for 100, 1000 hours and one year (8760 hours) 
of observation time with the ORT and 100 hours of WSRT and LOFAR. 
Auger data points reproduced from ~\cite{SemikozAuger} 
on the CR flux are shown on left side for comparison.}
\label{ortcrlimit}
\end{figure}

\begin{figure}[t]
\hbox{\hspace{0cm}
\hbox{\includegraphics[scale=1.3]{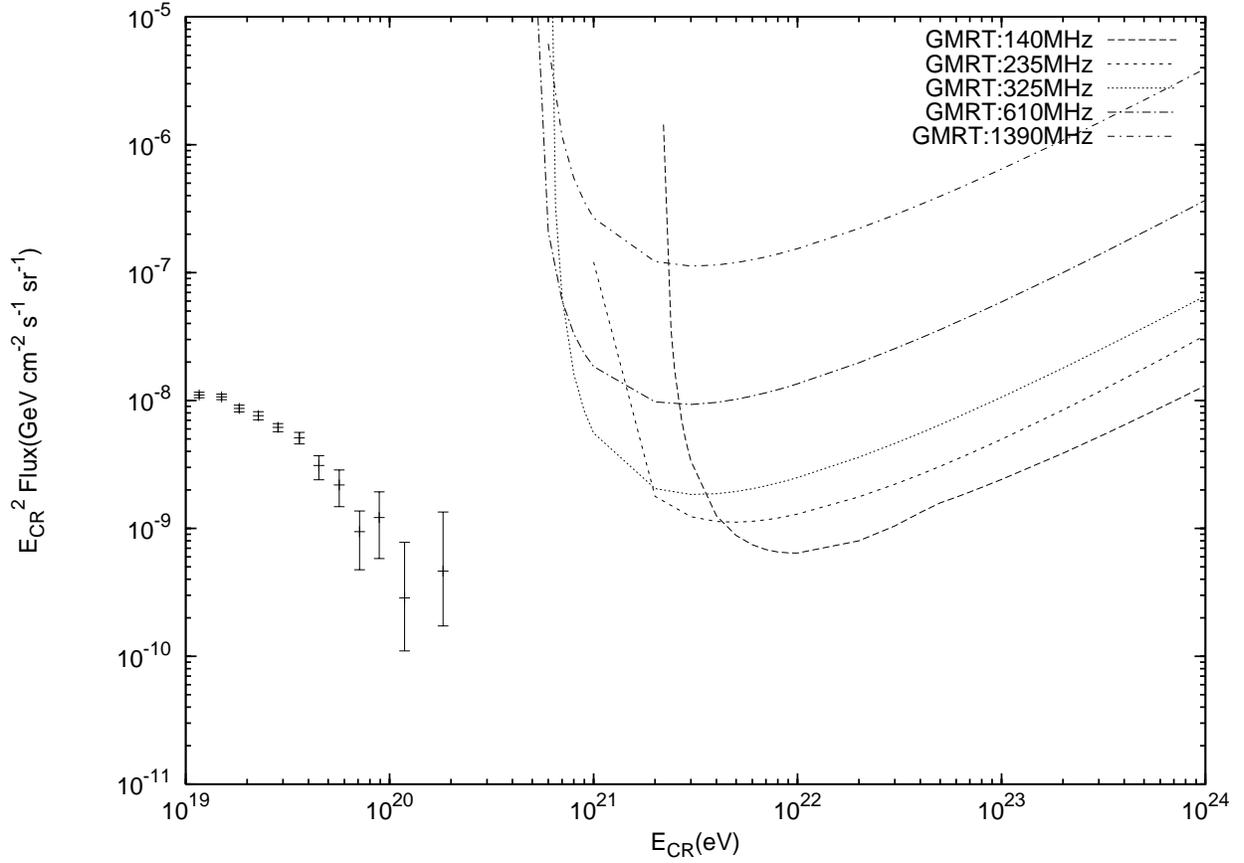}}}
\caption{Model independent limits on UHECR flux at different
frequencies for 100 hours of observation time with 
the GMRT for an incoherent array. Auger data points 
reproduced from ~\cite{SemikozAuger} on the 
CR flux are shown on left side for comparison.}
\label{crlimit100}
\end{figure}

\begin{figure}[t]
\hbox{\hspace{0cm}
\hbox{\includegraphics[scale=1.3]{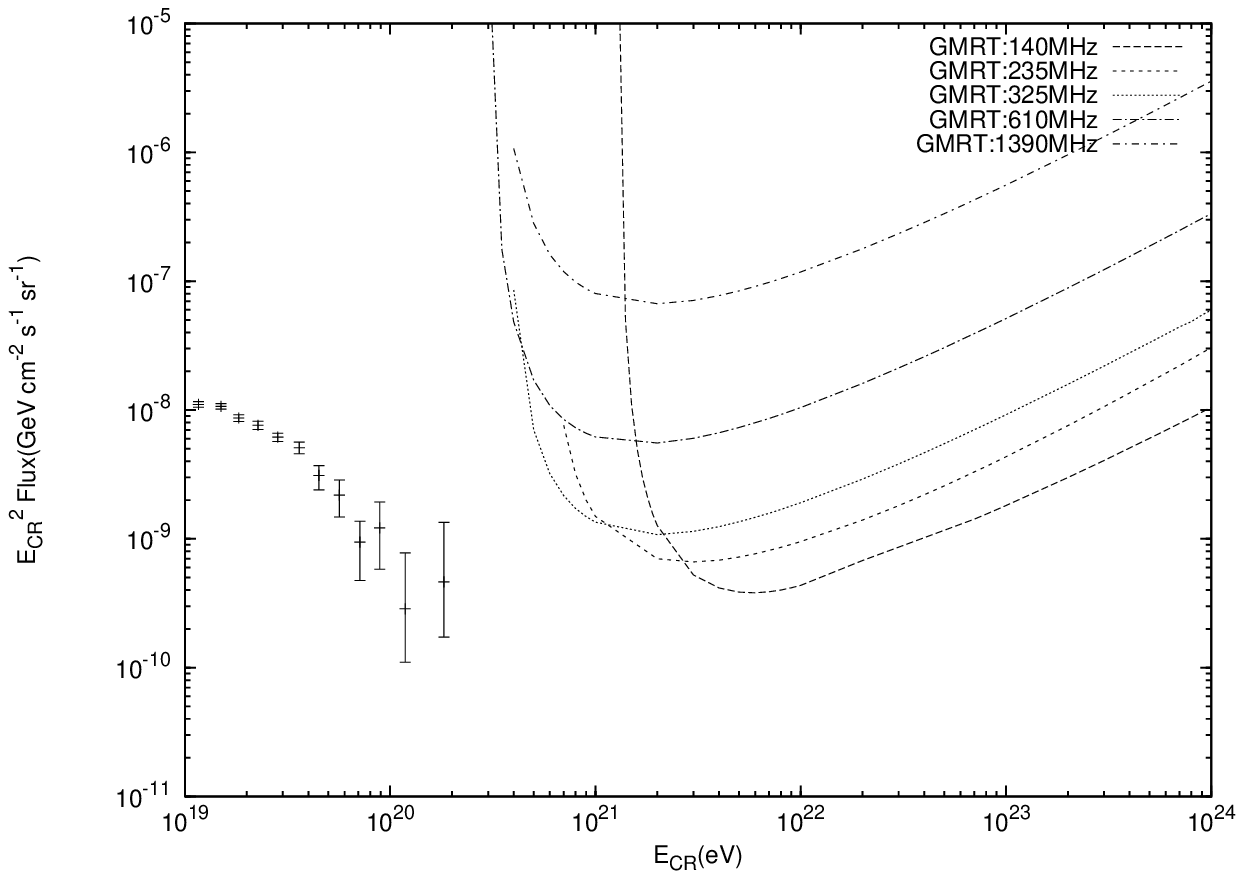}}}
\caption{Model independent limits on UHECR flux at different
frequencies for 100 hours of observation time with 
the GMRT for the 25 beams case. Auger data points 
reproduced from ~\cite{SemikozAuger} on the 
CR flux are shown on left side for comparison.}
\label{crblimit100}
\end{figure}

If no events are observed at GMRT and ORT over a time $T$ 
then an upper limit can be derived on UHE CR and neutrino 
fluxes at the Moon. 
The conventional model-independent limit ~\cite{FORTE} is given by
\begin{equation}
E_{i}^2 \frac{\mathrm{d}\Phi_{i}}{\mathrm{d}E_{\nu}} 
\leq s_\mathrm{up} \frac{E_
{i}}{A_{i}(E_s=y_i E_{i})T},
\end{equation}                  
where still $i=\left\{\nu,\mathrm{CR}\right\}$, $y_{CR}=1$ 
and $y_\nu=0.25$. The Poisson factor $s_\mathrm{up}=2.3$ 
for a limit at $90 \%$ confidence level. In Fig.\ref{ortcrlimit},
are shown prospective limits on the flux of the UHE CRs for
T=100, 1000, 8760 hours (one year) of the observation time with ORT.
Plots for WSRT and LOFAR for T=100 hours of the 
observation time are also shown. In 
Figs \ref{crlimit100} and \ref{crblimit100} are given 
model independent limits on UHE CR flux at different
frequencies of the GMRT for an incoherent array and 25 beams
case respectively for 100 hours of observations.

\begin{figure}[t]
\hbox{\hspace{0cm}
\hbox{\includegraphics[scale=1.3]{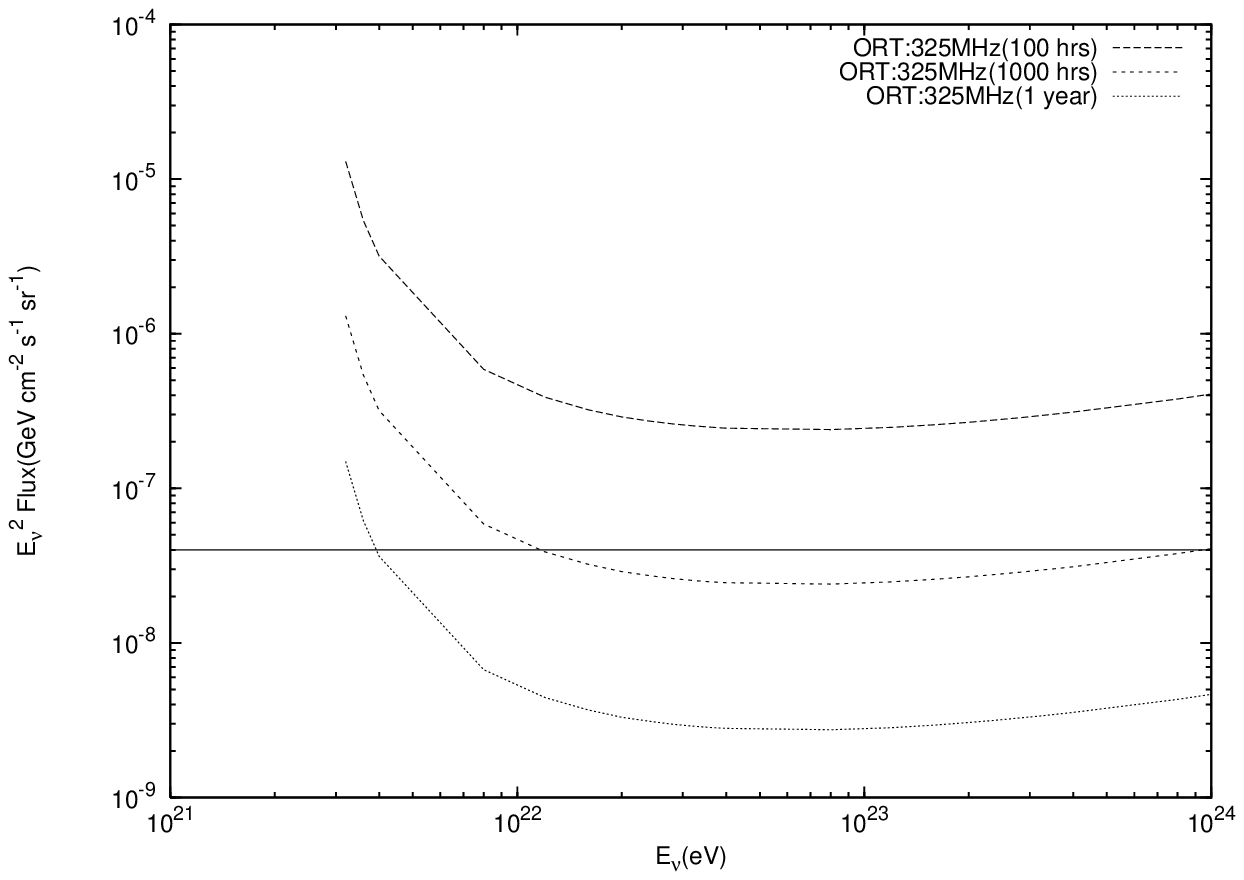}}}
\caption{Model independent limits on UHE neutrino flux 
at different frequencies for 100, 1000 hours and one year of observation 
time with the ORT. The solid horizontal line refers to 
the Waxman-Bahcall limit \cite{Waxman:1998yy}.}
\label{ortnulimit}
\end{figure}

\begin{figure}[t]
\hbox{\hspace{0cm}
\hbox{\includegraphics[scale=1.3]{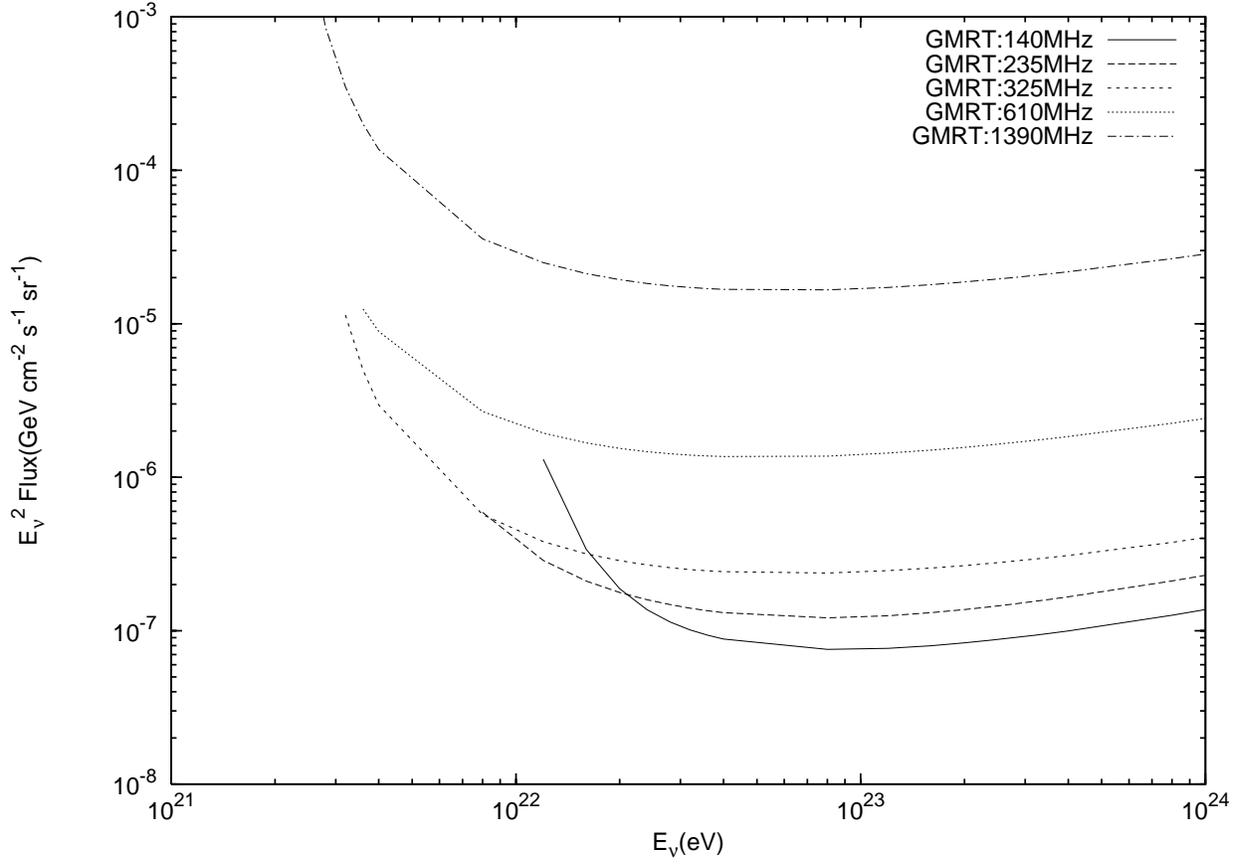}}}
\caption{Model independent limits on UHE neutrino flux
at different frequencies for 100 hours of observation time 
with the GMRT for an incoherent array.}
\label{nlimit100}
\end{figure}

\begin{figure}[t]
\hbox{\hspace{0cm}
\hbox{\includegraphics[scale=1.3]{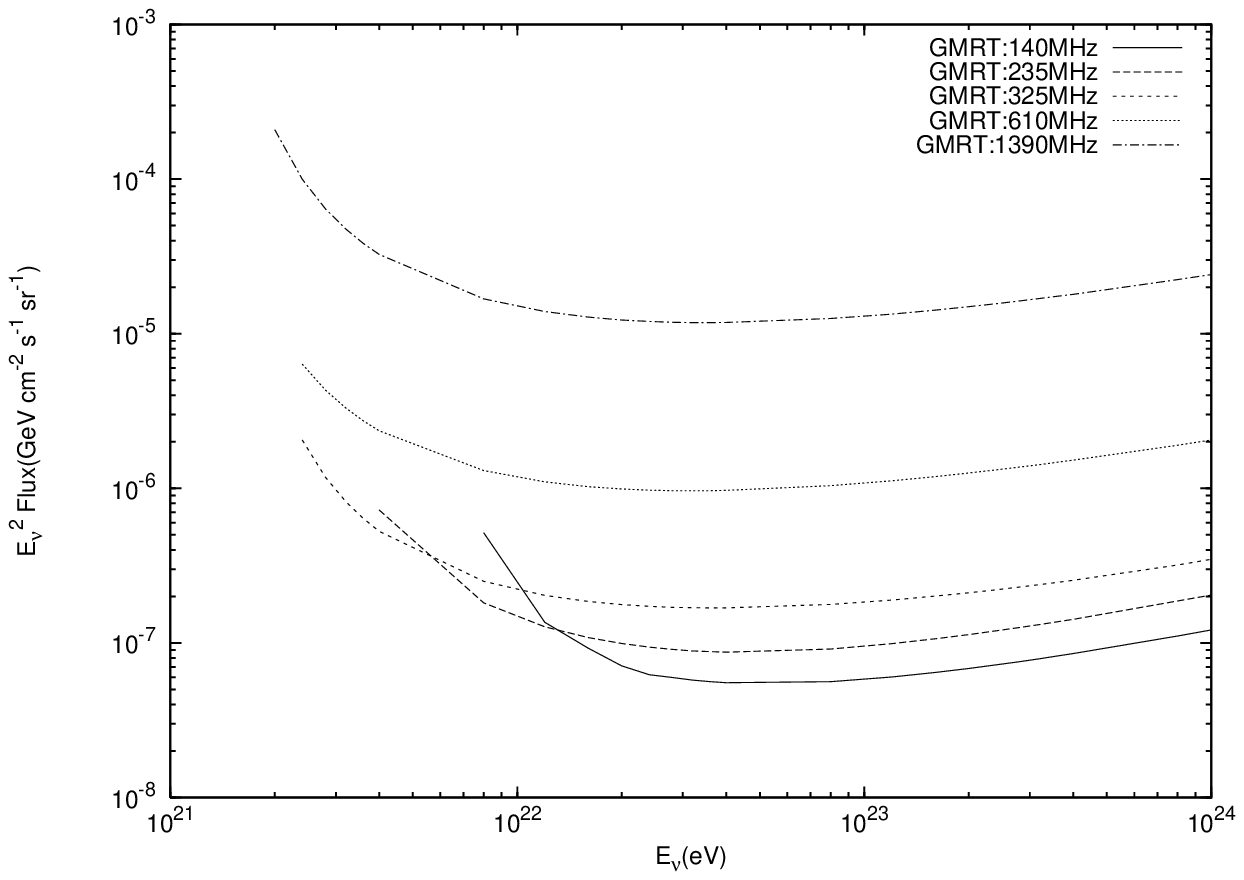}}}
\caption{Model independent limits on UHE neutrino flux
at different frequencies for 100 hours of observation time
with the GMRT for the 25 beams case.}
\label{nblimit100}
\end{figure}

Similarly for the UHE neutrinos, prospective limits on their 
flux for T=100, 1000 and 8760 hours of observation with ORT are given in
Fig. \ref{ortnulimit}. Figs \ref{nlimit100} and \ref{nblimit100} give limits
on the UHE neutrinos at different frequencies of the 
GMRT for an incoherent array and 25 beams case respectively 
for 100 hours of observations. For all our calculations 
we take $\sigma=25$ \cite{Scholten}.

It is clear from the plots that that low frequency observations give more 
stringent limits on the flux at the expense of a higher 
threshold. This is due to the well-known increase in the 
aperture ~\cite{Scholten, panda} from radiation 
spreading at lower frequencies.
Since many radio experiments exist for UHE neutrino detection,
we have compiled a comparison in Fig.\ref{fluxgmrt}. This figure
contains, the predicted thresholds of the ORT at 
325 MHz for 1 year of observation time, 
of the GMRTB (25 beams case) at 140 MHz for
100 hrs and 30 days of observation time and the already existing limits 
from RICE ~\cite{rice}, GLUE ~\cite{glue}, FORTE ~\cite{FORTE} and 
ANITA-lite ~\cite{anita}. Also we have indicated the prospective 
future limits that has been calculated for ANITA ~\cite{anita}, 
LOFAR~\cite{Scholten} or LORD ~\cite{gusev}. James and Protheroe
~\cite{james} have recently calculated sensitivity of the 
next generation of lunar Cerenkov observations for UHE CR and neutrinos.

In addition to search for UHE CR and neutrinos, simultaneous observations 
with the full array of the GMRT will provide radio maps of the Moon 
as a function of the lunar phase, giving 
information about the average thermal and electrical conductivity 
of the Moon’s regolith up to depth 
of $\sim$ 30 m to 100 m.  
Therefore, for the two experiments to be carried out simultaneously, 
it may be possible to get 2$\times$ 50 hours of observations in two GMRT 
Time Allocation Cycles. Also, observations with the ORT at the same time
will allow discrimination against man made RFI transients. 

\section{Discussions and Conclusion}  
It will be prudent to use both the ORT and the GMRT for 
searching for the UHE neutrinos. The new software correlator 
being installed at the GMRT
will allow forming $\sim$ 25 beams at 140 MHz to cover the Moon
at $\sim$ 140 MHz providing 2 bands of 16 MHz and $A_{eff} \sim$ 14250 $m^2.$
One may also
conveniently use the incoherent mode of the 
GMRT with $A_{eff} \sim$ 5203 $m^2.$
Although ORT with $A_{eff}=$8000 $m^2$ operates only 
at 325 MHz, it is well suited to track the Moon for 
hundreds of hours. The RFI is also much lower at Ooty 
than at the GMRT site. By using the new digital system 
installed recently at Ooty by Prabhu {\it et al}~\cite{prabhu} of the 
Raman Research Institute, in conjunction with the 
12 beams of the analogue system ORT and also it's upgrade,
it should be 
possible to reach adequate sensitivity to test 
the Waxman Bahcall limit proposed on theoretical 
arguments on the UHE particle flux. 

Proposed observations, 
particularly with the ORT will also provide arcsec resolutions
for celestial radio sources occulted by the Moon, and may 
also detect any transient celestial sources present 
in the antenna beam outside the disc of the Moon. Search for
UHE neutrinos will also allow simultaneous observations for
making radio maps of the Moon as a function of the lunar phase 
(full Moon, 5 and 15 days earlier and later), providing information
about the average thermal and electrical conductivity of the Moon's
regolith up to a depth of $\sim$ 30 m.  

\begin{figure}[t]
\hbox{\hspace{0cm}
\hbox{\includegraphics[scale=1.3]{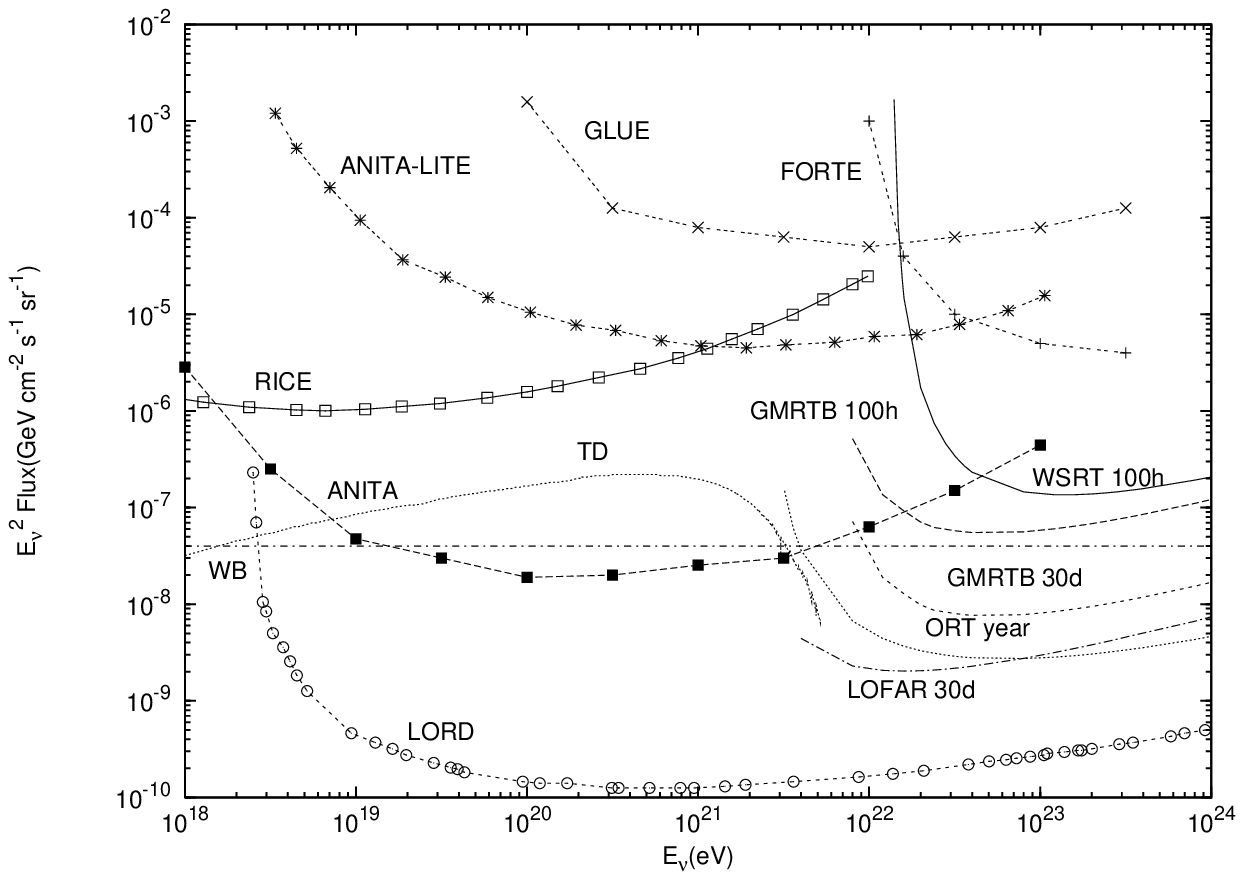}}}
\caption{Prospective flux limits on UHE neutrinos 
from GMRTB (25 beams), ORT, WSRT and LOFAR are shown for 
effective exposure times of $100$ hour
s, $30$ days, $1000$ hours and $1$ year (8760hours). 
The current best limits from 
radio experiments ANITA-lite \cite{anita}, GLUE \cite{glue}, 
FORTE \cite{FORTE}, and RICE \cite{rice} are shown. 
For comparison the expected limits from 
future experiments ANITA \cite{anita} and 
LORD \cite{gusev} are also included. The 
dotted horizontal WB line indicates the theoretical 
upper limit of Waxman-Bahcall 
\cite{Waxman:1998yy} on the cosmogenic neutrino flux. 
TD refers to the Topological Defect model with mass 
$M_X=2 \times 10^{22}$ eV described in \cite{td}.}
\label{fluxgmrt}
\end{figure}

The existence of UHE Neutrinos of
$> 10^{20}$ eV is implied by the 
detection of for $\sim 10^{19}$ eV. The extremely
high luminosity of the star burst galaxies, AGNs,
gamma ray burst are likely to accelerate protons to very 
high energies that get scattered by the CMBR photons 
producing a flux of UHE neutrinos. There are also predictions 
of their occurrence by more exotic sources in the early universe.
As may be seen from Fig. \ref{fluxgmrt}, observations with the ORT 
and GMRT will provide a threshold sensitivity of $\sim 10^{-8},$ 
being comparable to the current searches being 
made by other investigators. Detection of the UHE CR and neutrinos 
of $> 10^{20}$ eV would be of great importance for testing theories
of high energy physics and for understanding several phenomena 
of cosmological and astrophysical importance.


{\bf{Acknowledgement}} We thank T. Prabhu of the Raman Research Institute, 
P.K. Manoharan and A.J. Selvanayagam of the Radio Astronomy Centre Ooty 
and S. Sirothia of NCRA, Pune for many valuable discussions. The work of 
S.P. was supported by the Ministerio de Educacion y Ciencia
under Proyecto Nacional FPA2006-01105, and also by the Comunidad de Madrid
under Proyecto HEPHACOS, Ayuda de I+D S-0505/ESP-0346.

\end{document}